\documentclass[aps,prd,onecolumn,notitlepage,letterpaper,10pt]{revtex4-1}
\usepackage{style}

\begin{document}

\title{Ponderomotive Effects of Ultralight Dark Matter}

\author{Kevin Zhou\,\orcidlink{0000-0002-9810-3977}}
\email{kzhou7@berkeley.edu}
\affiliation{Berkeley Center for Theoretical Physics, University of California, Berkeley, CA 94720, USA}
\affiliation{Theoretical Physics Group, Lawrence Berkeley National Laboratory, Berkeley, CA 94720, USA}

\begin{abstract}
\noindent
I exhibit a new class of quadratic effects of ultralight dark matter. Axions, dark photons, and dilatons can exert rapidly oscillating forces, torques, and mass shifts on Standard Model particles. These effects average to zero at first order, but shift particle properties at second order, in analogy to the ponderomotive force in optics. Remarkably, these effects scale with the square of the amplitude of the dark matter field, even when the field's direct physical effects depend only on its derivatives. I calculate the resulting observables in electron $g_e - 2$ experiments using classical mechanics, recovering results previously derived using field theory. When considered properly, these particular experiments do not beat astrophysical bounds, but other precision experiments may have interesting sensitivity. 
\end{abstract}

\maketitle

{
\vspace{-10mm}
\hypersetup{linkcolor=black}
\tableofcontents
}

\begin{center}
\textit{Conventions and Notation}
\end{center}

We use a mostly-negative spacetime metric and natural units, $\hbar = c = 1$, with rationalized units for electromagnetic fields (i.e., SI units with $\epsilon_0 = \mu_0 = 1$). The electron has mass $m_e$ and charge $q < 0$. Angle brackets denote time averages over the dark matter oscillation timescale; observables are decomposed into slowly and quickly varying components by $\mathcal{O} = \langle \mathcal{O} \rangle + \Delta \mathcal{O}$. For brevity, we define the dimensionless combination $\lambda = \rhodm / (\mdm^2 m_e^2)$. 

\newpage
\section{Introduction}
\label{sec:intro}

Several recent works~\cite{Evans:2023uxh,Arza:2023wou,Evans:2024dty} have argued that in the presence of ultralight dark matter with dimensionless coupling $g$, observables in electron $g_e - 2$ experiments can be shifted by a fractional amount proportional to $g^2 \rhodm / (\mdm^2 m_e^2) \equiv g^2 \lambda$. Ref.~\cite{Evans:2023uxh} considered the dark photon, Ref.~\cite{Arza:2023wou} considered the dilaton and axion-fermion couplings, and Ref.~\cite{Evans:2024dty} considered axion-fermion and axion-photon couplings simultaneously. Though these effects are quadratic in a small coupling $g$, they are enhanced by $1/\mdm^2$ and therefore appear to be extremely strong in the ultralight limit. 

In Refs.~\cite{Evans:2023uxh,Arza:2023wou,Evans:2024dty}, these shifts were computed using Feynman diagrams with a background-corrected dark matter propagator, and applying techniques developed for thermal field theory~\cite{Donoghue:1984zz}. These calculations are technically involved, but physically opaque. In particular, axions and dark photons decouple in the limit $\mdm \to 0$, so we would expect their leading physical effects to scale with their time derivative, proportional to $\sqrt{\rhodm}$, but $\lambda$ instead scales as $(\sqrt{\rhodm} / \mdm)^2$, the square of the amplitude. However, analogous effects exist even in classical electromagnetism. In plasma physics, it is well-known that when an electron is placed in an electromagnetic wave with high angular frequency $\omega$ and electric field amplitude $\v{E}_0(\v{r})$, it experiences a time-averaged ``ponderomotive'' force~\cite{boot1958containment,PhysRevD.1.2738,zangwill2013modern}
\begin{equation} \label{eq:ordinary_ponderomotive}
\langle \v{F} \rangle = - \frac{q^2}{4 \, m_e \, \omega^2} \nabla(|\v{E}_0|^2).
\end{equation}
This force displays the same quadratic field dependence and $1/\omega^2$ enhancement. 

I will show, using elementary classical arguments, that dark photon, dilaton, and axion dark matter give rise to analogous ``dark ponderomotive'' effects with the same scaling. Crucially, they only exist when the dark matter field oscillates rapidly compared to a relevant experimental timescale, $\mdm \gg \omega_0$. The claims of strong sensitivity in Refs.~\cite{Evans:2023uxh,Arza:2023wou,Evans:2024dty} are incorrect because they assume these effects continue to strengthen in the regime $\mdm \ll \omega_0$, where the dark matter instead decouples.

For concreteness, we will focus on the leading electron $g_e - 2$ experiment~\cite{Fan:2022eto,Fan:2022oyb}, in which a nonrelativistic electron of mass $m_e$ performs cyclotron motion in a uniform magnetic field $\v{B}$ with angular frequency $\omega_c \simeq q B / m_e$, while its spin precesses at angular frequency $\omega_s \simeq (g_e / 2) (q B / m_e)$. In this experiment, the characteristic frequency of the electron's motion is $\omega_0 \sim \omega_c \sim \omega_s \sim \meV$, and the difference $\omega_s - \omega_c$ is measured. We will compute the shifts in the time-averaged cyclotron and spin precession frequencies. For illustrative purposes, we will also compute the shift in the electron mass, defined by the electron's time-averaged energy at rest, as this quantity is particularly simple.

We will begin with the dark photon in Sec.~\ref{sec:dark_photon}, as this case is most closely analogous to electromagnetism. Neglecting $\mathcal{O}(\vdm)$ effects, it acts solely via an oscillating force proportional to the kinetic mixing $\epsilon$, which we will see yields
\begin{equation}
\frac{\delta m_e}{m_e} \simeq \frac12 q^2 \epsilon^2 \lambda, \qquad \frac{\delta \omega_c}{\omega_c} \simeq - \frac56 q^2 \epsilon^2 \lambda, \qquad \frac{\delta \omega_s}{\omega_s} \simeq - \frac12 q^2 \epsilon^2 \lambda. 
\end{equation}
Along the way, we will explicitly see why these effects only exist when $\mdm \gg \omega_0$, and how the effect of the dark photon smoothly shuts off in the limit $\mdm \to 0$. 

In Sec.~\ref{sec:scalar}, we consider a dilaton $\phi$ with a Yukawa coupling $g_s$ to the electron. Here the leading effect is a variation of the electron mass proportional to $\phi$, which does not directly contribute to $\delta m_e$ because it averages to zero, but does shift $\omega_c$ and $\omega_s$. We will also see that the leading contribution to the mass shift $\delta m_e$ appears at $\mathcal{O}(\vdm^2)$, giving 
\begin{equation}
\frac{\delta m_e}{m_e} \simeq \frac32 g_s^2 \lambda \vdm^2, \qquad \frac{\delta \omega_c}{\omega_c} \simeq g_s^2 \lambda, \qquad \frac{\delta \omega_s}{\omega_s} \simeq g_s^2 \lambda.
\end{equation}
In Sec.~\ref{sec:axion} we consider the dimensionless axion-fermion coupling $g_d$ and axion-photon coupling $\bar{g}_\gamma$. We will review how the derivative and pseudoscalar forms of the axion-fermion coupling are equivalent, and show that the axion's sole leading effect depends on the cross term,
\begin{equation}
\frac{\delta \omega_s}{\omega_s} \simeq g_d \bar{g}_\gamma \lambda. 
\end{equation}
Finally, in Sec.~\ref{sec:conclusion} we compare to existing literature, and discuss experimental implications. 

\section{Dark Photon}
\label{sec:dark_photon}

In vacuum, the dark photon field ${A'}^\mu$, with associated field strength ${F'}^{\mu\nu}$, obeys the equation of motion 
\begin{equation}
\del_\mu {F'}^{\mu\nu} + \mdm^2 {A'}^\nu = 0.
\end{equation}
As a result, the modes of the dark photon field have mass $\mdm$, so each time derivative of the field yields a factor of $\mdm$, while each spatial gradient yields a factor of $\mdm \vdm$, where $\vdm \ll 1$ is the nonrelativistic dark matter velocity. The field obeys the condition $\del_\mu {A'}^\mu = 0$, which implies $A^0 \sim \vdm A^i$, so that the dark electric field is 
\begin{equation}
\v{E}' = - \frac{\del \v{A}'}{\del t} - \bm{\nabla} {A'}^0 = - \frac{\del \v{A}'}{\del t} + \mathcal{O}(\vdm^2)
\end{equation}
while the dark magnetic field $\v{B}' = \bm{\nabla} \times \v{A}'$ is itself $\mathcal{O}(\vdm)$. The energy density is 
\begin{equation}
\rhodm = \frac12 (|\v{E}'|^2 + |\v{B}'|^2) + \frac12 \mdm^2 (|\v{A}'|^2 - ({A'}^0)^2) = \frac12 |\v{E}'|^2 + \frac12 \mdm^2 |\v{A}'|^2 + \mathcal{O}(\vdm^2).
\end{equation}
Clear reviews of this material are given, e.g.~in Refs.~\cite{Caputo:2021eaa,Fedderke:2021aqo}. Throughout this section, we will neglect the dark matter velocity for simplicity. As a result, polarized dark photon dark matter is described by 
\begin{equation}
\v{A}' = \v{A}_0 \cos(\mdm t), \qquad |\v{A}_0| = \frac{\sqrt{2 \rhodm}}{\mdm}
\end{equation}
with $\v{E}' = - \del \v{A}' / \del t$. For simplicity, we will assume the dark photon is unpolarized, which implies that
\begin{equation} \label{eq:A_dir_avg}
\langle A'_i A'_j \rangle = \frac{\rhodm}{3 \mdm^2} \, \delta_{ij}.
\end{equation}
The dark photon interacts with the electron through a small kinetic mixing $\epsilon$, so that in the mass basis, the electron experiences an oscillating force, which leads to an oscillating shifts in its momentum. Defining $\v{F} = \langle \v{F} \rangle + \Delta \v{F}$ and $\v{p} = \langle \v{p} \rangle + \Delta \v{p}$, the leading contributions to these shifts are 
\begin{equation} \label{eq:dp_shift}
\Delta \v{F} \simeq q \epsilon \v{E}', \qquad \Delta \v{p} \simeq - q \epsilon \v{A}'.
\end{equation}
In a real $g_e - 2$ experiment, the dark photon will be partially shielded by the trap, but we neglect this for simplicity. We review how the electron responds to forces and torques in Sec.~\ref{sec:force_torque}, and compute ponderomotive effects in Sec.~\ref{sec:dp_ponderomotive}. More technical derivations are deferred to Sec.~\ref{sec:dp_alternative}, which can be skipped without loss of continuity. 

\subsection{Review: Relativistic Electrons}
\label{sec:force_torque}

It will turn out that, even though the ponderomotive effects we will derive apply for a nonrelativistic electron, some of these effects can only be found by considering the leading relativistic corrections to the electron's dynamics. We therefore begin by reviewing the equations of motion for a classical relativistic electron in an electromagnetic field. 

First, the force on the electron is 
\begin{equation} \label{eq:em_force}
\v{F} = \frac{d\v{p}}{dt} = q (\v{E} + \v{v} \times \v{B})
\end{equation}
where $\v{p} = \gamma m_e \v{v}$, which implies that in a uniform magnetic field, an electron orbits with the cyclotron angular frequency $\omega_c = q B / (\gamma m_e)$. As for the spin, in the nonrelativistic limit it obeys the equation of motion
\begin{equation} \label{eq:nr_spin}
\frac{d\v{S}}{dt} = \bm{\mu} \times \v{B} = \frac{q g_e}{2 m_e} \, \v{S} \times \v{B}
\end{equation}
where $g_e$ is the electron $g$-factor. In the relativistic case, the Bargmann--Michel--Telegdi equation~\cite{berestetskii1982quantum,jackson1999classical} governs the evolution of the spin four-vector $S^\mu$, equal to $(0, \v{S})$ in the electron's rest frame. In terms of $\v{S}$, it reads 
\begin{equation} \label{eq:rel_spin}
\frac{d\v{S}}{dt} = \v{S} \times \left( \frac{q g_e}{2m_e} \left( \v{B} - \frac{\gamma}{\gamma + 1} (\v{v} \cdot \v{B}) \v{v} - \v{v} \times \v{E} \right) + \frac{\gamma^2}{\gamma + 1} \v{v} \times \v{a} \right).
\end{equation}
Here, the field in parentheses is simply $\v{B}'/\gamma$, where $\v{B}'$ is the magnetic field in the electron's frame, and the final term is the Thomas precession, containing the additional Wigner rotation due to the electron's acceleration. If the electron's acceleration is governed by~\eqref{eq:em_force}, then~\eqref{eq:rel_spin} reduces to Thomas' equation, 
\begin{equation}
\frac{d\v{S}}{dt} = \frac{q}{m_e} \, \v{S} \times \left( \left( \frac{g_e}{2} - 1 + \frac{1}{\gamma} \right) \v{B} - \left( \frac{g_e}{2} - 1 \right) \frac{\gamma}{\gamma + 1} (\v{v} \cdot \v{B}) \v{v} -\left( \frac{g_e}{2} - \frac{\gamma}{\gamma + 1} \right) \v{v} \times \v{E} \right).
\end{equation}
Thus, the spin of an electron in a uniform magnetic field with $\v{v} \cdot \v{B} = 0$ will precess with angular frequency 
\begin{equation}
\omega_s = \omega_c + \frac{qB}{m_e} \frac{g_e-2}{2}.
\end{equation}
This result is the basis for experimental probes of $g_e - 2$. Since $g_e - 2$ is small, we will set $g_e = 2$ for simplicity throughout this work, which simplifies Thomas' equation to 
\begin{equation} \label{eq:thomas_g2}
\frac{d\v{S}}{dt} = \frac{q}{\gamma m_e} \, \v{S} \times \left( \v{B} - \frac{\gamma}{\gamma + 1} \, \v{v} \times \v{E}\right).
\end{equation}

\subsection{Ponderomotive Effects}
\label{sec:dp_ponderomotive}

Now we are prepared to calculate ponderomotive effects. First, the electron mass is defined by its time-averaged energy at rest. In the presence of the dark photon, an electron that is on average at rest will ``jitter'' in place, giving it a nonzero average kinetic energy. At leading order, its total energy is
\begin{equation} \label{eq:basic_mass_shift}
m_e + \delta m_e = \langle \gamma m_e \rangle \simeq m_e + \frac{\langle |\Delta \v{p}|^2 \rangle}{2 m_e}.
\end{equation}
Thus, dark photon dark matter induces a fractional mass shift of 
\begin{equation} \label{eq:dp_mass_shift}
\frac{\delta m_e}{m_e} \simeq \frac{\langle |\Delta \v{p}|^2 \rangle}{2 m_e^2} = \frac{q^2 \epsilon^2 \rhodm}{2 \mdm^2 m_e^2} = \frac12 q^2 \epsilon^2 \lambda. 
\end{equation}
This type of effect is well-known in optics, as it also occurs when an electron is placed in a rapidly oscillating electric field. As reviewed in Ref.~\cite{Heinzl:2022lly}, it was introduced in the 1960s to describe intensity-dependent effects in Compton scattering with laser pulses. A classical derivation of the mass shift, generalized to arbitrary pulse shapes, was given in Ref.~\cite{Eberly:1968zz}, and Ref.~\cite{PhysRevE.77.036402} contains a modern derivation which applies for more general fields. More recently, Refs.~\cite{Harvey:2012ie,Kohlfurst:2013ura} discussed prospects for directly measuring the mass shift in electron beams. 

As for the cyclotron frequency, it is changed because the dark photon modifies the average momentum an electron has, for a given average velocity. To see this, we write the velocity with its first relativistic correction, 
\begin{equation} \label{eq:vp_relation}
\v{v} = \frac{\v{p}}{\gamma m_e} \simeq \frac{\v{p}}{m_e} - \frac12 \frac{|\v{p}|^2 \, \v{p}}{m_e^3}.
\end{equation}
Taking the time average, terms with an odd number of powers of $\Delta \v{p}$ average to zero, so 
\begin{equation}
\langle \v{v} \rangle \simeq \frac{\langle \v{p} \rangle}{m_e} - \frac{1}{2 m_e^3} \left( |\langle \v{p} \rangle|^2 \langle \v{p} \rangle + \langle |\Delta \v{p}|^2 \rangle \langle \v{p} \rangle + 2 \langle (\Delta \v{p} \cdot \langle \v{p} \rangle) \Delta \v{p} \rangle \right).
\end{equation}
The final term can be simplified using~\eqref{eq:A_dir_avg}, giving
\begin{equation}
\langle (\Delta \v{p} \cdot \langle \v{p} \rangle) \Delta \v{p} \rangle = \frac13 \langle |\Delta \v{p}|^2 \rangle \langle \v{p} \rangle.
\end{equation}
Then in total, we have 
\begin{equation} \label{eq:vdp_final}
\langle \v{v} \rangle \simeq \left(1 - \frac12 \frac{|\langle \v{p} \rangle|^2}{m_e^2} - \frac{5}{6} \frac{\langle |\Delta \v{p}|^2 \rangle}{m_e^2} \right) \frac{\langle \v{p} \rangle}{m_e}
\end{equation}
The cyclotron frequency is proportional to the right-hand side. The $|\langle \v{p} \rangle|^2$ term gives the leading relativistic correction to the cyclotron frequency, while the $\langle |\Delta \v{p}|^2 \rangle$ term due to the dark photon gives a fractional shift of 
\begin{equation} \label{eq:dp_wc_shift}
\frac{\delta \omega_c}{\omega_c} \simeq -\frac{5}{6} q^2 \epsilon^2 \lambda.
\end{equation}
This result can also be derived by explicitly considering all forces on the electron, as we will see below. 

Finally, we consider the spin precession frequency for an electron at rest on average, $\v{v} = \Delta \v{v}$. The fastest method is to note that, since the dark electric field acts analogously to an ordinary electric field, Thomas' equation~\eqref{eq:thomas_g2} still holds if we replace $\v{E}$ with $\epsilon \v{E}'$, so that 
\begin{equation} 
\frac{d\v{S}}{dt} = \frac{q}{\gamma m_e} \, \v{S} \times \left( \v{B} - \frac{\gamma}{\gamma + 1} \, \Delta \v{v} \times (\epsilon \v{E}') \right).
\end{equation}
At leading order in $\Delta \v{v}$, we have $\Delta \v{v} \simeq \Delta \v{p}/m_e \propto \v{A}'$, which is $\pi/2$ out of phase with $\v{E}'$. Then the time average of the second term vanishes, and we simply have $\omega_s \simeq \langle q B / (\gamma m_e) \rangle$ where $1/\gamma \simeq 1 - |\Delta \v{v}|^2/2$, which gives the shift
\begin{equation} \label{eq:dp_ws_shift}
\frac{\delta \omega_s}{\omega_s} \simeq -\frac12 q^2 \epsilon^2 \lambda.
\end{equation}
It is also possible to derive this directly from the Bargmann--Michel--Telegdi equation, as we will see below. 

These three ponderomotive effects were previously derived in Ref.~\cite{Evans:2023uxh} using thermal field theory. That work's results for $\delta m_e$ and $\delta \omega_c$ agree with ours, but its $\delta \omega_s$ differs by a factor of $2$. However, the numeric coefficients of that work were originally derived in the ultrarelativistic limit, and may not be correct in the nonrelativistic limit~\cite{private}. 

In a real $g_e - 2$ experiment, the observed shifts would differ from the ideal results computed here, for several reasons. First, we have neglected shielding, which can strongly suppress the dark photon's effect~\cite{Chaudhuri:2014dla}, since it depends on the details of the experiment's geometry. Second, we have assumed that the dark photon is unpolarized, so that its effect on the electron is isotropic. However, dark photon dark matter can also be produced in a polarized state~\cite{Caputo:2021eaa}, in which case $\delta \omega_c$ would depend on the relative orientation of the polarization and the plane of the electron's orbit. There are also subleading orientation-dependent effects due to the dark matter's average velocity, which we have neglected. This would give rise, for instance, to a dark analogue of the ordinary ponderomotive force~\eqref{eq:ordinary_ponderomotive}. 

Regardless, the derivations here illustrate how it is possible to have effects that scale with $\lambda$. The force due to the dark photon is proportional to $\sqrt{\rhodm}$, but its oscillation period is proportional to $1/\mdm$. When $\mdm \gg \omega_0$, the electron is effectively free over this oscillation timescale, over which it accumulates a velocity of amplitude $\Delta v \propto \sqrt{\rhodm} / \mdm$. All of the effects above depend on $(\Delta v)^2 \propto \rhodm / \mdm^2$. By contrast, when $\mdm \lesssim \omega_0$, the electron would be able to build up this speed. For instance, in the limit $\mdm \ll \omega_0$, the dark photon exerts an effectively constant force on the electron. This shifts the electron's position in the trap by $\sim \sqrt{\rhodm} / k$, where the coefficient $k$ quantifies the trap's stiffness. As the dark matter oscillates, this small shift varies adiabatically, inducing an electron velocity scaling as $\mdm \sqrt{\rhodm}$. Thus, in the limit $\mdm \to 0$, the dark photon has no observable effect, as expected. 

\subsection{Alternative Derivations}
\label{sec:dp_alternative}

To further justify the above results, we can also derive $\delta \omega_c$ and $\delta \omega_s$ by explicitly considering how the electron responds to forces and torques. First, taking the time derivative of~\eqref{eq:vp_relation} gives an acceleration of 
\begin{equation} \label{eq:accel_expr}
\v{a} \simeq \frac{\v{F}}{m_e} - \frac{1}{2 m_e^3} (|\v{p}|^2 \v{F} + 2 (\v{p} \cdot \v{F}) \v{p}).
\end{equation}
We now decompose $\v{F} = \langle \v{F} \rangle + \Delta \v{F}$ and $\v{p} = \langle \v{p} \rangle + \Delta \v{p}$ and compute the time-averaged acceleration $\langle \v{a} \rangle$. For the quantity in parentheses in~\eqref{eq:accel_expr}, terms with an odd number of factors of $\Delta \v{F}$ and $\Delta \v{p}$ time-average to zero. The ponderomotive effect comes from the terms with two factors, which are
\begin{equation} \label{eq:pf_corrections}
\left\langle |\v{p}|^2 \v{F} + 2 (\v{p} \cdot \v{F}) \v{p} \right\rangle \supset \left\langle (\Delta \v{p} \cdot \Delta \v{p}) \langle \v{F} \rangle + 2 (\Delta \v{p} \cdot \langle \v{F} \rangle) \Delta \v{p} + 2 (\Delta \v{p} \cdot \langle \v{p} \rangle) \Delta \v{F} + 2 (\Delta \v{p} \cdot \Delta \v{F}) \langle \v{p} \rangle + 2 (\langle\v{p}\rangle \cdot \Delta \v{F}) \Delta \v{p} \right\rangle.
\end{equation}
We know the dominant contribution to the oscillating force is $\Delta \v{F} \supset q \epsilon \v{E}'$, which implies $\Delta F / \langle F \rangle \sim (\mdm/\omega_c) \Delta p / \langle p \rangle \gg \Delta p / \langle p \rangle$. Therefore, we would expect the last three terms above, which contain $\Delta \v{F}$, to dominate. 

However, it turns out that the time averages of these terms sum to zero. Since $\v{E}'$ and $\v{A}'$ are $\pi/2$ out of phase with each other, we have $\langle E'_i A'_j \rangle = 0$, which implies that if $\Delta \v{F} = q \epsilon \v{E}'$, then $\langle \Delta p_i \, \Delta F_j \rangle = 0$ and each term individually averages to zero. Now, there is also a subdominant part of $\Delta \v{F}$ due to the force from the background magnetic field, 
\begin{equation} \label{eq:Delta_F_expr}
\Delta \v{F} = q \epsilon \v{E}' + q \Delta \v{v} \times \v{B} \simeq q \epsilon \v{E}' + \frac{q}{m_e} \Delta \v{p} \times \v{B}.
\end{equation}
It is smaller than the direct dark photon force by $\omega_c / \mdm$, but it must be accounted for, since $\Delta F / F$ started out parametrically larger by $\mdm / \omega_c$. With this contribution included, the fourth term in~\eqref{eq:pf_corrections} still averages to zero, because $\langle \Delta \v{p} \cdot \Delta \v{F} \rangle = 0$. For an unpolarized dark photon, the third and fifth terms are nonzero, but cancel each other. 

We are thus left with the first two terms of~\eqref{eq:pf_corrections}. Simplifying using~\eqref{eq:A_dir_avg} again, and using the fact that $\langle \v{p} \rangle \cdot \langle \v{F} \rangle = 0$ because $\langle \v{F} \rangle = q \langle \v{v} \rangle \times \v{B}$, we conclude that 
\begin{equation} 
\langle \v{a} \rangle \simeq \frac{\langle \v{F} \rangle}{m_e} \left( 1 - \frac12 \frac{|\langle \v{p} \rangle|^2}{m_e^2} - \frac{5}{6} \frac{\langle |\Delta \v{p}|^2 \rangle}{m_e^2} \right)
\end{equation}
in agreement with~\eqref{eq:vdp_final}. The cyclotron frequency is proportional to $\langle \v{a} \rangle$, giving the same $\delta \omega_c$ as found before. 

To describe spin precession, we take~\eqref{eq:rel_spin} with its first relativistic correction, with $\v{E} \to \epsilon \v{E}'$ and $g_e = 2$. Within the correction terms, we can work to leading order in $\v{v} = \Delta \v{v}$, which implies $\Delta \v{v} \simeq \Delta \v{p} / m_e$ and $\v{a} \simeq \Delta \v{F} / m_e$, so that
\begin{equation} 
\frac{d\v{S}}{dt} = \v{S} \times \bm{\Omega}, \qquad \bm{\Omega} \simeq \frac{q}{m_e} \left( \v{B} - \frac{1}{2 m_e^2} (\Delta \v{p} \cdot \v{B}) \Delta \v{p} - \frac{\epsilon}{m_e} \Delta \v{p} \times \v{E}' \right) + \frac{1}{2 m_e^2} \, \Delta \v{p} \times \Delta \v{F}.
\end{equation}
The spin precession rate is proportional to $\langle \bm{\Omega} \rangle$. As we have argued above, $\langle \Delta \v{p} \times \v{E}' \rangle$ vanishes, and using~\eqref{eq:Delta_F_expr} gives
\begin{equation}
\langle \bm{\Omega} \rangle \simeq \frac{q}{m_e} \left(\v{B} - \frac{1}{2 m_e^2} \left\langle(\Delta \v{p} \cdot \v{B}) \Delta \v{p} + (\Delta \v{p} \times \v{B}) \times \Delta \v{p} \right\rangle \right)
\end{equation}
Using~\eqref{eq:A_dir_avg}, the terms in brackets simplify to $1/3$ and $2/3$ times $\langle |\Delta \v{p}|^2 \rangle \, \v{B}$, respectively, recovering the $\delta \omega_s$ found before. We also learn that $2/3$ of the shift is due to Thomas precession. 

\section{Dilaton}
\label{sec:scalar}

Next, we consider a scalar field $\phi$ coupled to the electron mass. This corresponds to a Lagrangian
\begin{equation}
\mathcal{L} = \bar{\Psi} (i \slashed{\del} - (m_e + g_s \phi)) \Psi
\end{equation}
where $\Psi$ is the electron Dirac field. For a field mode of momentum $\v{k}$ and amplitude $\phi_0$, the average energy density is 
\begin{equation}
\langle \rho \rangle = \left\langle \frac12 \dot{\phi}^2 + \frac12 \mdm^2 \phi^2 + \frac12 |\bm{\nabla} \phi|^2 \right\rangle = \frac{\phi_0^2}{4} (\omega^2 + \mdm^2 + |\v{k}|^2) = \frac{\phi_0^2}{2} (\mdm^2 + |\v{k}|^2).
\end{equation}
The dark matter velocity distribution consists of an isotropic component shifted by the relative velocity of the laboratory and the galactic halo. We drop this ``dark matter wind'' effect for simplicity. We define $\vdm$ so that the weighted average of $|\v{k}|^2$ over all modes is $\mdm^2 \vdm^2$. Then the amplitude of the total field obeys, to order $\vdm^2$,
\begin{equation} \label{eq:moving_field_amp}
\rhodm \simeq \mdm^2 (1 + \vdm^2) \langle \phi^2 \rangle \simeq \langle \dot{\phi}^2 \rangle.
\end{equation}
In Sec.~\ref{sec:dilaton_uniform}, we will warm up by computing the $\mathcal{O}(g_s^2 \lambda)$ ponderomotive effects, where we can neglect $\vdm$ and take $\phi$ to be spatially uniform. In Sec.~\ref{sec:dilaton_ham}, we derive the nonrelativistic Hamiltonian for the electron accounting for spatial gradients of $\phi$. In Sec.~\ref{sec:dilaton_vde}, we compute the associated forces and torques, and find the $\mathcal{O}(g_s^2 \lambda \vdm^2)$ mass shift. 

\subsection{Uniform Field Effects}
\label{sec:dilaton_uniform}

By symmetry, a spatially uniform dilaton $\phi$ cannot directly affect the momentum or spin of an electron. However, it does lead to a time-dependent shift in the electron mass, so that all instances of $m_e$ in the equations of motion of Sec.~\ref{sec:force_torque} should be replaced with 
\begin{equation}
\bar{m}_e = m_e + g_s \phi.
\end{equation}
To see this explicitly, we can derive the electron's nonrelativistic Hamiltonian using Pauli elimination, as was done for the axion-fermion coupling in Ref.~\cite{Berlin:2023ubt}. The four-component electron field obeys the equation of motion
\begin{equation}
(i \slashed{\del} - m_e - g_s \phi - q \slashed{A}) \Psi = 0
\end{equation}
where we have set the scalar potential to zero, since we are only interested in situations with a uniform magnetic field. We define two-component nonrelativistic spinors by 
\begin{equation} \label{eq:psi_defs}
\Psi = e^{- i (m_e + g_s \phi) t} \begin{pmatrix} \psi \\ \tilde{\psi} \end{pmatrix}
\end{equation}
which, in the Dirac representation of the gamma matrices, obey 
\begin{align}
i \del_t \, \psi &= \bm{\pi} \cdot \bm{\sigma} \, \tilde{\psi}, \label{eq:psi_eom} \\
(i \del_t + 2 m_e + 2 g_s \phi) \, \tilde{\psi} &= \bm{\pi} \cdot \bm{\sigma} \, \psi, \label{eq:psit_eom}
\end{align}
where we defined the kinetic momentum $\bm{\pi} = \v{p} - q \v{A}$. We solve~\eqref{eq:psit_eom} for $\tilde{\psi}$ at leading order in the nonrelativistic expansion, plug the result back into~\eqref{eq:psi_eom}, and define $i \del_t \psi \equiv H \psi$ for
\begin{equation} \label{eq:uniform_ham}
H = \frac{(\bm{\pi} \cdot \bm{\sigma})^2}{2 \bar{m}_e} = \frac{(\v{p} - q \v{A})^2 - q \v{B} \cdot \bm{\sigma}}{2 \bar{m}_e}.
\end{equation}
This is simply the Pauli Hamiltonian with $m_e$ replaced with $\bar{m}_e$, as anticipated. 

Since the dilaton field averages to zero, this effect produces no mass shift on average, $\langle \bar{m}_e \rangle = m_e$, at any order in $g_s$. However, at second order it produces shifts to the cyclotron and spin precession frequencies, because these quantities depend on $1/\bar{m}_e$. For example, for an electron at rest, the equation of motion~\eqref{eq:nr_spin} for the spin becomes 
\begin{equation} \label{eq:dilaton_spin_eq_uniform}
\frac{d \v{S}}{dt} = \frac{q}{m_e + g_s \phi} \, \v{S} \times \v{B}
\end{equation}
which corresponds to an average spin precession frequency of 
\begin{equation} \label{eq:dilaton_avg_ws}
\omega_s = \left\langle \frac{q B}{m_e + g_s \phi} \right\rangle \simeq \frac{qB}{m_e} \left( 1 + \frac{g_s^2 \langle \phi^2 \rangle}{m_e^2} \right).
\end{equation}
In addition, whenever the spin precesses, it has an oscillating component with an amplitude proportional to $g_s$. 

As for forces, the dilaton modifies the kinetic momentum to $\bm{\pi} = (m_e + g_s \phi) \v{v}$ but does not impose an oscillating force, $\Delta \v{F} = 0$. As a result, the acceleration is
\begin{equation} \label{eq:accel_dilaton}
\v{a} = \frac{1}{m_e + g_s \phi} \left( \v{F} - \frac{g_s \dot{\phi} \, \v{p}}{m_e + g_s \phi} \right).
\end{equation}
Taking the time average, noting that $\langle \phi \rangle = 0$, $\langle \phi \dot{\phi} \rangle = 0$, and $\v{F} = \langle \v{F} \rangle$, gives an increased acceleration of 
\begin{equation} \label{eq:dilaton_avg_a}
\langle \v{a} \rangle = \frac{\langle \v{F} \rangle}{m_e} \left( 1 + \frac{g_s^2 \langle \phi^2 \rangle}{m_e^2} \right).
\end{equation}
In addition, whenever the electron moves, its velocity has an oscillating component proportional to $g_s \langle \v{v} \rangle$. From~\eqref{eq:dilaton_avg_ws} and~\eqref{eq:dilaton_avg_a}, we conclude that the cyclotron and spin precession frequencies are shifted by 
\begin{equation}
\frac{\delta \omega_c}{\omega_c} = \frac{\delta \omega_s}{\omega_s} \simeq g_s^2 \lambda,
\end{equation}
in agreement with Ref.~\cite{Arza:2023wou}.

\subsection{Full Nonrelativistic Hamiltonian}
\label{sec:dilaton_ham}

Now we will calculate the nonrelativistic Hamiltonian, accounting for the leading corrections due to gradients of $\phi$. For brevity, we temporarily drop subscripts on $g_s$ and $m_e$. The final result is~\eqref{eq:reduced_dilaton_ham}.

Since $\phi$ is no longer spatially homogeneous, it is unnatural to define nonrelativistic spinors as in~\eqref{eq:psi_defs}. Instead, it is more convenient to only factor out a phase of $e^{- i m t}$, so that $\psi$ and $\tilde{\psi}$ obey 
\begin{align}
(i \del_t - g \phi) \, \psi &= \bm{\pi} \cdot \bm{\sigma} \, \tilde{\psi}, \label{eq:psi_eom_2} \\
(i \del_t + 2 m + g \phi) \, \tilde{\psi} &= \bm{\pi} \cdot \bm{\sigma} \, \psi. \label{eq:psit_eom_2}
\end{align}
To make power counting explicit, we will keep all powers of $g$ in the numerator for now. By comparison with~\eqref{eq:uniform_ham}, the Pauli Hamiltonian itself is $\mathcal{O}(1/m)$, and the relevant linear and quadratic dilaton terms above appear at $\mathcal{O}(g / m^2)$ and $\mathcal{O}(g^2 / m^3)$ respectively. We will therefore keep terms out to these orders. Fine structure corrections appear at $\mathcal{O}(1/m^3)$, but we will suppress them because they are irrelevant for the leading dilaton effects. 

To begin, solving~\eqref{eq:psit_eom_2} for $\tilde{\psi}$ approximately gives 
\begin{align} \label{eq:psi_iter}
\tilde{\psi} &= \frac{\bm{\pi} \cdot \bm{\sigma}}{2 m} \, \psi - \frac{i \del_t + g \phi}{2m} \, \tilde{\psi} \\
&= \left(\frac{\Pi}{2 m} - \frac{i \del_t + g \phi}{2 m} \left(\frac{\Pi}{2 m} - \frac{i \del_t + g \phi}{2 m} \frac{\Pi}{2 m} \right) \right) \psi + \mathcal{O}(1/m^4) \label{eq:psi_final}
\end{align} 
where we iterated the equation in the second step, and defined $\Pi = \bm{\pi} \cdot \bm{\sigma}$. Plugging this back into~\eqref{eq:psi_eom_2} gives
\begin{equation}
i \del_t \psi \simeq g \phi \, \psi + \frac{\Pi^2}{2 m} \, \psi - \Pi \left(\frac{i \del_t + g \phi}{2 m} \left(\frac{\Pi}{2 m} - \frac{i \del_t + g \phi}{2 m} \frac{\Pi}{2 m} \right) \right) \psi.
\end{equation}
First, let's extract the $\mathcal{O}(g/m^2)$ terms. The inner factor of $i \del_t + g \phi$ is irrelevant, as it leads to terms which are suppressed by too many powers of $1/m$. As for the outer factor of $i \del_t + g \phi$, the time derivative can act on $\psi$, which yields $g \phi \, \psi$ at leading order in $1/m$. We therefore have 
\begin{equation}
i \del_t \psi \supset - \frac{g}{4 m^2} \, \Pi (\phi \Pi + \Pi \phi).
\end{equation}
As for the $\mathcal{O}(g^2/m^3)$ terms, similar reasoning gives 
\begin{equation}
i \del_t \psi \supset \frac{1}{8 m^3} \, \Pi (i \del_t + g \phi)^2 \Pi \, \psi \simeq \frac{g^2}{8 m^3} (\Pi \phi^2 \Pi + \Pi^2 \phi^2 + 2 \Pi \phi \Pi \phi).
\end{equation}
At this point, we can define a fiducial Hamiltonian $\bar{H}$ by defining $i \del_t \psi = \bar{H} \psi$. 

This Hamiltonian is not quite Hermitian, because the equations of motion preserve the magnitude of $\Psi$ rather than $\psi$. As in Appendix E of Ref.~\cite{Berlin:2023ubt}, we define a renormalized two-component wavefunction $\psi_{\text{nr}}$ whose norm is preserved,
\begin{equation} \label{eq:norm_equality}
\int d^3\v{x} \, \Psi^\dagger \Psi \equiv \int d^3 \v{x} \, \psi_{\text{nr}}^\dagger \psi_{\text{nr}}
\end{equation}
If $\tilde{\psi} = \mathcal{S} \psi$, then~\eqref{eq:norm_equality} holds if we take 
\begin{equation}
\psi_{\text{nr}} = M \psi \simeq \left( 1 + \frac{\mathcal{S}^\dagger \mathcal{S}}{2} \right) \psi.
\end{equation}
Reading off $\mathcal{S}$ from~\eqref{eq:psi_final}, and keeping only terms out to $\mathcal{O}(g/m^3)$, we find
\begin{equation}
M \simeq 1 + \frac{\Pi^2}{8 m^3} - \frac{g}{8 m^3} \left( \Pi \phi \Pi + \frac12 \phi \Pi^2 + \frac12 \Pi^2 \phi \right).
\end{equation}
The proper nonrelativistic Hamiltonian is $H = \bar{H} + [M, \bar{H}] + i \del_t M$, which contains fine structure corrections which we suppress, and additional $\mathcal{O}(g/m^2)$ and $\mathcal{O}(g^2/m^3)$ terms due to the commutator of $M$ with the $g \phi$ term in $\bar{H}$. 

The resulting Hamiltonian is 
\begin{equation} \label{eq:final_dilaton_ham}
H = g \phi + \frac{\Pi^2}{2 m} - \frac{g}{8 m^2} \, (\Pi^2 \phi + 2 \Pi \phi \Pi + \phi \Pi^2) + \frac{g^2}{8 m^3} \left( \Pi \phi^2 \Pi + \frac12 \Pi^2 \phi^2 + \frac12 \phi^2 \Pi^2 + \Pi \phi \Pi \phi + \phi \Pi \phi \Pi \right) 
\end{equation}
which is manifestly Hermitian. As a check, note that when $\phi$ is spatially uniform, it commutes with $\Pi$, so that
\begin{equation}
H = g \phi + \left( 1 - \frac{g \phi}{m} + \frac{g^2 \phi^2}{m^2} \right) \frac{\Pi^2}{2 m} = g \phi + \frac{(\bm{\pi} \cdot \bm{\sigma})^2}{2 (m + g \phi)} + \mathcal{O}(g^3/m^4).
\end{equation}
This contains the Hamiltonian found in Sec.~\ref{sec:dilaton_uniform}, along with an explicit $g\phi$ term which reflects the dilaton-induced shift in energy for an electron at rest. 

More generally, we can evaluate~\eqref{eq:final_dilaton_ham} to find the terms that depend on gradients of $\phi$. The $\mathcal{O}(g/m^2)$ terms yield 
\begin{equation} \label{eq:linear_dilaton_ham}
H \supset \frac{g}{8 m^2} \, \nabla^2 \phi - \frac{g}{2m^2} \, \phi (\bm{\pi} \cdot \bm{\sigma})^2 + \frac{ig}{2 m^2} \, (\bm{\nabla} \phi \cdot \bm{\sigma})(\bm{\pi} \cdot \bm{\sigma})
\end{equation}
where we simplified using $(\bm{\sigma} \cdot \bm{\nabla})(\bm{\sigma} \cdot \bm{\nabla}) \phi = \nabla^2 \phi$. Similarly, the $\mathcal{O}(g^2/m^3)$ terms yield
\begin{equation} \label{eq:quad_dilaton_ham}
H \supset \frac{g^2}{2 m^3} \left( \phi^2 (\bm{\pi} \cdot \bm{\sigma})^2 - \frac12 \phi \nabla^2 \phi - \frac12 |\bm{\nabla} \phi|^2 - 2 i \, (\phi \bm{\nabla} \phi \cdot \bm{\sigma}) \, (\bm{\pi} \cdot \bm{\sigma}) \right).
\end{equation}
Here we are interested in $\mathcal{O}(g^2 \vdm^2)$ ponderomotive effects, so not all contributions in the Hamiltonian will be relevant. In particular, the terms above with two gradients acting on $\phi$ yield forces which in turn depend on their own gradients, and are therefore $\mathcal{O}(\vdm^3)$. Furthermore, we can drop the final $\mathcal{O}(g^2)$ term in~\eqref{eq:quad_dilaton_ham} because it only yields force contributions which average to zero, so that its averaged effects only appear at even higher order in $g$. 

Therefore, dropping irrelevant terms and restoring subscripts, our final Hamiltonian is
\begin{equation} \label{eq:reduced_dilaton_ham}
H \simeq g_s \phi + \frac{1}{2 (m_e + g_s \phi)} \, (\bm{\pi} \cdot \bm{\sigma})^2 + \frac{i g_s}{2 m_e^2} \, (\bm{\nabla} \phi \cdot \bm{\sigma})(\bm{\pi} \cdot \bm{\sigma}).
\end{equation}
Note that it is slightly non-Hermitian, but this will not affect the calculations that follow. 

\subsection{Velocity-Dependent Effects}
\label{sec:dilaton_vde}

We now calculate torques and forces using the Heisenberg equation of motion, $d \mathcal{O} / dt = \del_t \mathcal{O} + i [H, \mathcal{O}]$. We always work in the classical limit, so quantum expectation values are implicit below. First, the spin $\v{S} = \bm{\sigma}/2$ evolves as 
\begin{equation} \label{eq:dilaton_spin_eq}
\frac{d\v{S}}{dt} \simeq \frac{q}{m_e + g_s \phi} \, \v{S} \times \v{B} - \frac{g_s}{m_e^2} ((\bm{\nabla} \phi \cdot \v{S}) \, \bm{\pi} - \bm{\nabla} \phi \, (\bm{\pi} \cdot \v{S})),
\end{equation}
where we used the identities $(\v{a} \cdot \bm{\sigma}) \bm{\sigma} = \v{a} + i \bm{\sigma} \times \v{a}$ and $(\bm{\sigma} \times \v{a}) (\bm{\sigma} \cdot \v{b}) = - \v{a} \times \v{b} + i ((\v{a} \cdot \v{b}) \bm{\sigma} - (\v{a} \cdot \bm{\sigma}) \v{b})$. This generalizes our earlier equation~\eqref{eq:dilaton_spin_eq_uniform} by adding a $\mathcal{O}(v_e \vdm)$ torque due to the dilaton gradient, which to my knowledge has not been previously pointed out, and may be of interest for precision experiments with relativistic electrons. 

Next, to compute the acceleration we first define the velocity operator by 
\begin{equation}
\v{v} = i [H, \v{x}] \simeq \frac{1}{m_e + g_s \phi} \, \bm{\pi} + \frac{i g_s}{2 m_e^2} (\bm{\nabla} \phi \cdot \bm{\sigma}) \bm{\sigma}.
\end{equation}
The corresponding acceleration $\v{a} = i [H, \v{v}] + \del \v{v}/\del t$ contains many terms. For simplicity, we keep only the leading $\mathcal{O}(g_s)$ terms, dropping terms which are suppressed by additional powers of $g_s$, $\vdm$, the electron velocity $\v{v}$, or the ratios $\mdm/m_e$ or $\omega_s / m_e$. The final result, for a uniform magnetic field, is 
\begin{equation} \label{eq:dilaton_accel_eq}
\v{a} \simeq \frac{1}{m_e + g_s \phi} \left(q \v{v} \times \v{B} - g_s (\bm{\nabla} \phi + \dot{\phi} \v{v}) \right)
\end{equation}
where the $\bm{\nabla} \phi$ term arises from the commutator of $\bm{\pi}$ with $H \supset g_s \phi$, and the $\dot{\phi}$ term arises from $\del \v{v} / \del t$. As expected, this reduces to~\eqref{eq:accel_dilaton} in the absence of spatial gradients. 

The equations of motion~\eqref{eq:dilaton_spin_eq} and~\eqref{eq:dilaton_accel_eq} can be used to compute $\mathcal{O}(g_s^2 \lambda \vdm^2)$ ponderomotive effects. In this work we will consider just the mass shift for simplicity. For an electron on average at rest, we have $\Delta \v{F} \simeq - g_s \bm{\nabla} \phi$. This leads to a nonzero average kinetic energy from oscillatory motion, as for the dark photon, which shifts the mass by
\begin{equation}
\frac{\delta m_e}{m_e} \supset \frac{\langle |\Delta \v{p}|^2 \rangle}{2 m_e^2} \simeq \frac{g_s^2}{2 m_e^2 \mdm^2} \, \langle |\bm{\nabla} \phi|^2 \rangle \simeq \frac12 g_s^2 \lambda \vdm^2.
\end{equation}
This is what was found in Ref.~\cite{Arza:2023wou}. However, there is another, subtler contribution to the mass shift, which occurs because the gradient force pushes the electron to regions with higher $\phi$, which causes $\langle \phi \rangle$ to become nonzero. We have 
\begin{equation}
\Delta \v{x} \simeq \frac{g_s}{m_e \mdm^2} \bm{\nabla} \phi
\end{equation}
where the sign is positive, because the electron's displacement is opposite in sign to its acceleration. This gives
\begin{equation}
\frac{\delta m_e}{m_e} \supset \frac{g_s \langle \phi \rangle}{m_e} \simeq \frac{g_s}{m_e} \, \langle \Delta \v{x} \cdot \bm{\nabla} \phi \rangle \simeq \frac{g_s^2}{m_e^2 \mdm^2} \, \langle |\bm{\nabla} \phi|^2 \rangle.
\end{equation}
Summing these effects gives a total mass shift of 
\begin{equation}
\frac{\delta m_e}{m_e} \simeq \frac32 g_s^2 \lambda \vdm^2.
\end{equation}

\section{Axion}
\label{sec:axion}

Finally, we consider the axion's derivative coupling $g_d$ and pseudoscalar coupling $g_p$ to the electron, and its coupling $\bar{g}_\gamma$ to the photon, 
\begin{equation}
\mathcal{L} = \frac{g_d}{2 m_e} (\del_\mu a) \, \bar{\Psi} \gamma^\mu \gamma^5 \Psi - g_p \, a \bar{\Psi} i \gamma^5 \Psi - \frac14 \frac{\bar{g}_\gamma}{m_e} \, a F \tilde{F}.
\end{equation}
Note that $g_d$ and $g_p$ are both related to their dimensionful forms by a factor of $2 m_e$, and the usual dimensionful axion-photon coupling is $g_\gamma = \bar{g}_\gamma / m_e$. Due to the number of couplings in play, we will neglect $\vdm$ corrections for simplicity, so that for an axion profile $a(t) = a_0 \sin(\mdm t)$, the amplitude obeys $\rhodm \simeq \mdm^2 a_0^2/2$. In the following subsections, we consider each coupling in turn. 

\subsection{The Derivative Coupling}

For the derivative coupling, it was shown in Ref.~\cite{Berlin:2023ubt} that the nonrelativistic Hamiltonian is 
\begin{equation} \label{eq:deriv_ham}
H \simeq \frac{(\bm{\pi} \cdot \bm{\sigma})^2}{2 m_e} - \frac{g_d}{2 m_e} \bm{\nabla} a \cdot \bm{\sigma} - \frac{g_d}{4 m_e^2} \left( \dot{a} (\bm{\pi} \cdot \bm{\sigma}) + (\bm{\pi} \cdot \bm{\sigma}) \dot{a} \right).
\end{equation}
As in Sec.~\ref{sec:dilaton_ham}, we could have worked to greater accuracy to capture $\mathcal{O}(g_d^2/m_e^3)$ terms in the Hamiltonian. However, since the axion is derivatively coupled, these terms are suppressed by additional time derivatives and spatial gradients, making them irrelevant even for computing $\mathcal{O}(\lambda \vdm^2)$ ponderomotive effects. The most significant such term is $H \supset g_d^2 \dot{a}^2 / 8 m_e^3$. It is a direct mass shift, but it has two time derivatives, and is thus suppressed by $\mdm^2$ relative to the analogous dilaton term. Thus, the effects identified in Sec.~\ref{sec:dilaton_uniform} are not relevant for the derivative coupling. 

Next, the force that follows from~\eqref{eq:deriv_ham}, in the presence of a uniform magnetic field $\v{B}$, is 
\begin{equation} \label{eq:derivative_force}
\v{F} \simeq q \v{v} \times \v{B} - \frac{g_d}{m_e} \ddot{a} \, \v{S}
\end{equation}
where we have discarded axion terms which are parametrically suppressed, relative to the term kept here, by factors of $\vdm$ or $\omega_s / \mdm$. Compared to the dark photon case~\eqref{eq:dp_shift}, the leading axion-induced force is suppressed by $\mdm/m_e$, so that none of the effects considered in Sec.~\ref{sec:dark_photon} lead to relevant ponderomotive effects. 

Finally, the axion also induces an additional spin torque, 
\begin{equation} \label{eq:axion_fermion_torque}
\frac{d \v{S}}{dt} \simeq \frac{q}{m_e} \, \v{S} \times \v{B} + \frac{g_d}{m_e} \, \v{S} \times (\bm{\nabla} a + \dot{a} \v{v})
\end{equation}
where we have dropped terms suppressed by more powers of $\vdm$ or $v_e$. (The full expression for the spin torque, at all orders in $v_e$, is given in Ref.~\cite{Silenko:2021qgc}.) In principle, this torque could shift the spin precession frequency, either due to axion gradients or due to the velocity induced by the axion's force. 

However, it turns out that no such effects occur for the derivative coupling alone. First, consider the spin torque proportional to $\dot{a} \v{v}$. For an electron at rest on average, the force in~\eqref{eq:derivative_force} gives $\v{v} = \Delta \v{v} = - g_d \dot{a} \v{S} / m_e^2$. Comparing the magnitudes of the axion-induced ``effective'' magnetic field and the real magnetic field, we have
\begin{equation}
\frac{g_d \dot{a} v/q}{B} \sim g_d^2 \lambda \, \frac{\mdm^2}{q B}.
\end{equation}
In the high $\mdm$ limit this effect seems to be even more important than the $\mathcal{O}(\lambda)$ effects we have been encountering throughout this work. However, the cross product $\v{S} \times \Delta \v{v}$ for the axion-induced torque actually evaluates to zero. Moreover, we have $qB \simeq m_e \omega_c \sim (20 \, \eV)^2$, so that for this experiment, ultralight dark matter always has $\mdm^2 \ll qB$. 

As for the $\bm{\nabla} a$ term, it causes the orientation of the spin to jitter about its mean value, but it does not change the average spin precession frequency. This is easiest to see by considering the spin $\v{S}'$ in the frame rotating with angular frequency $\bm{\omega}_s = - q \v{B} / m_e$, which obeys
\begin{equation}
\frac{d \v{S}'}{dt} = \frac{g_d}{m_e} \, \v{S}' \times \bm{\nabla} a.
\end{equation}
Decomposing $\v{S}' = \langle \v{S}' \rangle + \Delta \v{S}'$, we see that components of $\bm{\nabla} a$ transverse to $\langle \v{S}' \rangle$ give rise to the oscillating part $\Delta \v{S}'$. For an isotropic dark matter velocity distribution, we have 
\begin{equation}
\langle |\Delta \v{S}'|^2 \rangle = \frac23 \frac{g_d^2 \vdm^2}{m_e^2} \, \langle a^2 \rangle \simeq \frac23 g_d^2 \lambda \vdm^2. 
\end{equation}
Since $\v{S}'$ has fixed magnitude, this leads to a reduction of the magnitude of the average spin $\langle \v{S}' \rangle$, 
\begin{equation} \label{eq:spin_mag_shift}
\frac{|\langle \v{S}' \rangle|}{|\v{S}'|} \simeq 1 - \frac13 g_d^2 \lambda \vdm^2.
\end{equation}
That is, the spin can never be fully aligned in a given direction. (Similar effects also occur for the dark photon and dilaton.) However, $d \langle \v{S}' \rangle / dt$ vanishes, so that the spin precession frequency is unchanged. By comparison, Ref.~\cite{Arza:2023wou} also found that no shifts appear at $\mathcal{O}(g_d^2 \lambda)$, but claims there is an $\mathcal{O}(g_d^2 \lambda \vdm^2)$ correction to the spin precession frequency. We will further discuss this discrepancy in the conclusion. 

\subsection{The Pseudoscalar Coupling}

The pseudoscalar coupling is almost equivalent to the derivative coupling. It was shown in Ref.~\cite{Berlin:2023ubt} (see also Ref.~\cite{Smith:2023htu}) that at linear order in the coupling, the resulting nonrelativistic Hamiltonian is equivalent to that of the derivative coupling, up to a unitary transformation. However, at quadratic order additional terms arise, which can shift the electron mass. To see this explicitly, note that the equations of motion for the pseudoscalar coupling are
\begin{align}
i \del_t \psi &= (\bm{\pi} \cdot \bm{\sigma} + i g_p a) \, \tilde{\psi}, \\
(i \del_t + 2 m_e) \tilde{\psi} &= (\bm{\pi} \cdot \bm{\sigma} - i g_p a) \, \psi.
\end{align}
Applying Pauli elimination, the nonrelativistic Hamiltonian contains the quadratic term
\begin{equation}
H \supset \frac{1}{2m_e} (\bm{\pi} \cdot \bm{\sigma} + i g_p a) (\bm{\pi} \cdot \bm{\sigma} - i g_p a) \supset \frac{g_p^2 a^2}{2 m_e}
\end{equation}
in addition to other, subdominant terms. This gives a mass shift of $\delta m_e / m_e \simeq g_p^2 \lambda / 2$, as was found in Ref.~\cite{Arza:2023wou}. 

However, this result is misleading, as the pseudoscalar coupling is not shift symmetric at quadratic order. More properly, one should start from the derivative coupling and apply a chiral field redefinition to the electron field. This eliminates the derivative coupling while applying a chiral rotation to the electron mass term, giving 
\begin{equation}
\mathcal{L} \supset \bar{\Psi} (i \slashed{\del} - e^{i g_d a \gamma^5 / m_e} m_e) \Psi = \bar{\Psi} (i \slashed{\del} - (i a g_d \gamma^5) - (m_e - g_d^2 a^2 / 2 m_e)) \Psi + \mathcal{O}(g_d^3)
\end{equation}
along with an anomaly-induced axion-photon coupling, which we neglect. The resulting Lagrangian contains the pseudoscalar coupling with $g_p = g_d$, but also an additional quadratic term which decreases the electron mass, cancelling the purported mass shift, and restoring agreement with the result for the derivative coupling. 

The more general lesson here is that quadratic effects of linear terms in the Lagrangian can contribute to observables at the same order as the leading effects of quadratic terms in the Lagrangian. To consistently compute either class of effects, it is important to consider the other. 

\subsection{The Photon Coupling}
\label{sec:axion_photon}

In the presence of a background electromagnetic field, the axion-photon coupling leads to additional source terms in Maxwell's equations, which produce axion-induced fields $\v{E}_a$ and $\v{B}_a$. For simplicity we neglect axion gradients, so that the only effect of the axion is to modify Ampere's law to
\begin{equation}
\bm{\nabla} \times \v{B} = \dot{\v{E}} + \v{J} + g_\gamma \dot{a} \v{B}.
\end{equation}
For an experiment with a uniform background magnetic field $B$ applied across a length scale $R$, it is well-known~\cite{Ouellet:2018nfr} that the electric and magnetic fields induced by the axion's ``effective current'' are suppressed in the limit $\mdm R \ll 1$, scaling as $E_a \sim (g_\gamma a B) (\mdm R)^2$ and $B_a \sim (g_\gamma a B) (\mdm R)$ respectively. This reflects the fact that the axion-photon coupling only depends on the axion's derivatives. 

However, for the experiment of Ref.~\cite{Fan:2022eto}, the electron trap has radius $R \sim \text{cm} \sim 10^{-5} \, \eV$. Thus, our assumption $\mdm \gg \omega_0$ automatically implies $\mdm R \gg 1$, and in this case we instead have $E_a \sim B_a \sim g_\gamma a B$. Both $\v{E}_a$ and $\v{B}_a$ contain contributions which vary in space on the small length scale $1/\mdm$, but spatially average to zero. We neglect these terms for simplicity, since their effects should average to zero for a generic electron trajectory. The spatial average of $\v{B}_a$ vanishes, so we are left with only the spatial average of $\v{E}_a$, 
\begin{equation} \label{eq:axion_e_field}
\v{E}_a \supset - g_\gamma a \v{B} = - \frac{\bar{g}_\gamma a \v{B}}{m_e}.
\end{equation}

The effect of the field in~\eqref{eq:axion_e_field} is small compared to that of the dark photon, because 
\begin{equation}
\frac{E_a}{\epsilon E'} \sim \frac{\bar{g}_\gamma}{\epsilon} \frac{B}{m_e \mdm} \sim \frac{\bar{g}_\gamma}{q \epsilon} \frac{\omega_c}{\mdm}.
\end{equation}
Thus, the axion-photon coupling does not induce ponderomotive effects at $\mathcal{O}(\bar{g}_\gamma^2 \lambda)$. However, it can produce an $\mathcal{O}(\bar{g}_\gamma g_d \lambda)$ effect in combination with the derivative coupling, because the latter yields forces and torques that depend on the electron's velocity. The leading velocity-dependent correction to the force, given in Ref.~\cite{Berlin:2023ubt}, is too suppressed, but the velocity-dependent torque in~\eqref{eq:axion_fermion_torque} is relevant. 

For concreteness, if we let $a(t) = a_0 \sin(\mdm t)$, then the electron gains an oscillating velocity 
\begin{equation}
\v{v} = \frac{q \bar{g}_\gamma \v{B}}{m_e^2 \mdm} \, a_0 \cos(\mdm t).
\end{equation}
The average spin torque is proportional to 
\begin{equation}
\left\langle \v{B} + \frac{g_p}{q} \dot{a} \v{v} \right\rangle \simeq \left\langle 1 + \frac{g_p \bar{g}_\gamma}{m_e^2} a_0^2 \cos^2(\mdm t) \right\rangle \v{B}
\end{equation}
which leads to an increased spin precession frequency of
\begin{equation}
\frac{\delta \omega_s}{\omega_s} = g_d \bar{g}_\gamma \lambda
\end{equation}
in agreement with Ref.~\cite{Evans:2024dty}. This shift is the leading ponderomotive effect of the axion. All other ponderomotive effects, which we do not compute here, are suppressed by powers of $\vdm^2$ or $v_e^2$.

\section{Discussion}
\label{sec:conclusion}

\subsection{Comparison to Prior Work}

In Refs.~\cite{Evans:2023uxh,Arza:2023wou,Evans:2024dty}, shifts of the electron mass, cyclotron frequency, and spin precession frequency were computed by considering tree-level Feynman diagrams with two external dark matter particles. Specifically, these works reinterpret those diagrams as loop diagrams with an internal dark matter line, whose propagator is modified by the dark matter background. They then compute the non-Lorentz invariant correction to the electron self-energy, infer an effective Hamiltonian, and simplify it with a Foldy--Wouthuysen transformation, after which the relevant shifts can be read off. We have seen that their results almost always agree parametrically with what we have found classically, though in a few cases the numeric coefficients differ. 

For the dark photon, some disagreement in numeric coefficients is to be expected, as Ref.~\cite{Evans:2023uxh} drops some relevant terms. In addition, for the $\mathcal{O}(\vdm^2)$ mass shift for the dilaton computed in Sec.~\ref{sec:dilaton_vde}, the difference might arise either because we account for more effects here, or due to an $\mathcal{O}(1)$ difference in the definition of $\vdm$. 

The most serious discrepancy is the parametric difference in $\delta \omega_s$ for the axion's derivative coupling. In~\eqref{eq:spin_mag_shift} we found that, though the time-averaged spin $\langle \v{S} \rangle$ has a reduced magnitude, $|\v{S}| \simeq (1 + g_d^2 \lambda \vdm^2/3) |\langle \v{S} \rangle|$, this averaged spin precesses with an unchanged frequency. By contrast, Refs.~\cite{Arza:2023wou,Evans:2024dty} find a shift of $\delta \omega_s / \omega_s \simeq g_d^2 \lambda \vdm^2/3$. 

I speculate that this discrepancy is due to their use of ``background-dependent spinors'', first introduced in Ref.~\cite{PhysRevD.28.340}, which affect the wavefunction renormalization. In our language, this could correspond to writing the Hamiltonian in terms of $\langle \v{S} \rangle$, without accounting for the fact that the spin torque still acts on the entire spin vector $\v{S}$. Interestingly, it was already noted in the foundational work Ref.~\cite{Donoghue:1984zz} that the choice of spinors affects the calculation of the electron spin precession frequency in a thermal photon background, with several papers finding contradictory results. Since that work, a number of additional prescriptions have been proposed~\cite{PhysRevD.53.4232,PhysRevD.55.6287,PhysRevD.58.105023,PhysRevD.63.083503}, which each yield physically distinct results. Reviews of Big Bang Nucleosynthesis, which takes place in a thermal background, have highlighted this issue as an ongoing controversy which contributes a small theoretical uncertainty~\cite{Lopez:1998vk,Serpico:2004gx,Iocco:2008va,Pitrou:2018cgg}. 

Regardless, this work shows that $\mathcal{O}(\lambda)$ effects can be computed classically. Refs.~\cite{Evans:2023uxh,Arza:2023wou,Evans:2024dty} claim that such effects are a result of Bose enhancement, which had been neglected in earlier works. However, it would be more correct to say that here, Bose enhancement is simply how the familiar classical fact that the effect of a field scales with its amplitude is equivalently described in quantum language. There is nothing qualitatively missing from existing calculations which treat ultralight dark matter as a classical field.

Our method is simple and physically transparent. In the field theoretic approach, many form factors appear, and it must be explicitly checked that the electron charge is not renormalized and that the electromagnetic current remains conserved; these facts are manifest in our classical approach. As we have discussed in Sec.~\ref{sec:dp_ponderomotive}, it is also manifest that ponderomotive effects only apply when $\mdm \gg \omega_0$. The field theoretic approach makes the same assumption, since it effectively integrates out the axion, but only implicitly.

We could have computed $\mathcal{O}(\lambda v_e^2)$ effects as well, but we neglected them since $v_e \ll \vdm \sim 10^{-3}$ in many precision experiments. However, bound electrons effectively have $v_e \sim \alpha \sim 10^{-2}$, and electrons in storage rings have $v_e \approx 1$. It would therefore be interesting to compute dark ponderomotive effects using a fully relativistic particle Lagrangian or Hamiltonian. Such calculations have been performed for point charges~\cite{PhysRevLett.75.4622} and Dirac particles~\cite{PhysRevA.92.062124} in strong, rapidly oscillating electromagnetic fields, and for particles in more general electromagnetic fields~\cite{PhysRevE.77.036402}. The case of dark matter could be handled similarly, and would likely be simpler because the dark matter is always weakly coupled. 

\subsection{Experimental Implications}

We have seen that for a generic dimensionless coupling $g$, an ultralight dark matter background can shift the electron $g_e - 2$ by $\sim g^2 \lambda = g^2 \rhodm / (m_e^2 \mdm^2)$. Refs.~\cite{Evans:2023uxh,Arza:2023wou,Evans:2024dty} state that constraints from $g_e - 2$ experiments become more powerful than astrophysical bounds roughly when $\mdm \lesssim 10^{-15} \, \eV$, and become arbitrarily strong in the limit $\mdm \to 0$. However, in reality these effects only exist when $\mdm \gg \omega_0 \sim \meV$. Since we numerically have $\lambda \simeq (3 \times 10^{-9} \, \eV / \mdm)^2$, in this regime ponderomotive effects are even weaker than the vacuum one-loop correction $\sim g^2 / (16 \pi^2)$, which in turn is much weaker than astrophysical bounds. 

However, ponderomotive effects could be relevant in precision experiments with characteristic frequencies $\omega_0 \lesssim 10^{-9} \, \eV \sim \text{MHz}$. For example, rapidly oscillating forces and torques acting on Standard Model particles could affect the gravitational force between pairs of such particles by $\mathcal{O}(g^2 \lambda)$. This could produce violations of the equivalence principle, visible in satellite, torsion pendulum, and atom interferometer experiments~\cite{Graham:2015ifn}. It would also be interesting to consider signatures in storage rings, though this requires generalizing to relativistic electrons. On the other hand, searches for ``variations of fundamental constants'', e.g.~with atomic clocks, are generally \textit{not} sensitive to ponderomotive effects. By definition, those experiments operate in the regime $\mdm \ll \omega_0$, while ponderomotive effects require $\mdm \gg \omega_0$. 

Since ponderomotive effects appear at zero frequency, searches for them would inherently have broadband sensitivity in $\mdm$. They could be distinguished from backgrounds through their dependence on the local dark matter field's amplitude, which varies over the coherence time $\sim 1 / (\mdm \vdm^2)$, and which for dark photons can be modulated by electromagnetic shielding. In addition, some ponderomotive effects would depend on the dark matter's relative velocity, which varies diurnally and annually. 

Since ponderomotive effects are second-order signatures of linear ultralight dark matter couplings, it is interesting to compare them to the leading signatures of quadratic couplings, proportional to the square of the field. Generically, we would expect them to be comparable, as they would both be $\mathcal{O}(1/\Lambda^2)$ for a new physics scale $\Lambda$. One important difference is that in the presence of these quadratic couplings, matter can locally affect the mass of the dark matter field. This can shield the field on Earth, in which case many precision experiments become ineffective~\cite{Banerjee:2022sqg}. Interestingly, shielding also produces a constant force due to reflection of the dark matter wind~\cite{Day:2023mkb}, though it differs parametrically from our ponderomotive effects, which appear even in the absence of a dark matter wind. Alternatively, the field can become unstable and grow to large values. For linear couplings, these complications do not appear. 

As an example, several recent works~\cite{Kim:2022ype,Banerjee:2023bjc,Flambaum:2023bnw,Beadle:2023flm,Kim:2023pvt,Bauer:2024hfv} have noted that since the QCD axion coupling $\sim (a/f_a) \, G^{\mu\nu} \tilde{G}_{\mu\nu}$ breaks the axion's shift symmetry, it induces $\mathcal{O}(1/f_a^2)$ quadratic couplings, such as $a^2 F^{\mu\nu} F_{\mu\nu}$ and $a^2 \bar{\Psi} \Psi$ for a nucleon or electron $\Psi$. These couplings can yield competitive constraints at low $\mdm$ because they are not suppressed by derivatives. However, working with these ``exceptionally light'' QCD axions comes at a cost. They require complex model building to avoid tuning, and are strongly constrained~\cite{Hook:2017psm,Balkin:2022qer,Gomez-Banon:2024oux,Kumamoto:2024wjd} because the axion can develop large field values inside stars, white dwarfs, and neutron stars. 

By contrast, we have shown in Sec.~\ref{sec:axion_photon} that an axion-like particle with only linear couplings to photons and electrons, both of which depend only on the axion's derivatives, can yield an observable effect proportional to $a^2$. This suggests that a variety of quadratic effects have yet to be found. 

\acknowledgments

I thank Ariel Arza, Itay Bloch, Jason Evans, Nicolas Rodd, and Samuel Wong for discussions. This work is supported by the Office of High Energy Physics of the U.S. Department of Energy under contract DE-AC02-05CH11231.

\newpage 
\bibliographystyle{utphys3}
\bibliography{Ponderomotive}

\end{document}